\documentclass[aps,prb,reprint,superscriptaddress,longbibliography]{revtex4-2}

\usepackage{amsmath}
\usepackage{amsfonts}
\usepackage{graphicx}
\usepackage{xspace}
\usepackage{subfigure}
\usepackage{hyperref}
\usepackage{enumitem}
\usepackage{xcolor}
\usepackage{comment}
\usepackage{xr}
\usepackage{siunitx}
\usepackage{float}
\DeclareSIUnit{\oersted}{Oe}

\usepackage[utf8]{inputenc}
\begin{document}

\title{Universal route towards field-free electrically polarity-reversible Josephson diode}

\author{Pramod. K. Sharma$^\dagger$}
\affiliation{Department of Metallurgical Engineering and Materials Science, Indian Institute of Technology Bombay, Powai, Mumbai 400076, India}
\author{Sagnik Banerjee$^\dagger$}
\affiliation{Department of Physics, Indian Institute of Technology Bombay, Powai, Mumbai 400076, India}
\author{Biswajit Dutta}
\affiliation{Department of Metallurgical Engineering and Materials Science, Indian Institute of Technology Bombay, Powai, Mumbai 400076, India}
\author{Vansh Singhal}
\affiliation{Department of Metallurgical Engineering and Materials Science, Indian Institute of Technology Bombay, Powai, Mumbai 400076, India}
\author{Pushpak Banerjee}
\affiliation{Department of Metallurgical Engineering and Materials Science, Indian Institute of Technology Bombay, Powai, Mumbai 400076, India}
\author{Sonam Bhakat}
\affiliation{Department of Metallurgical Engineering and Materials Science, Indian Institute of Technology Bombay, Powai, Mumbai 400076, India}
\author{Hridis K. Pal}
\affiliation{Department of Physics, Indian Institute of Technology Bombay, Powai, Mumbai 400076, India}
\affiliation{Centre of Excellence in Quantum Information, Computation, Science and Technology, Indian Institute of Technology Bombay, Powai, Mumbai 400076, India}
\author{Avradeep Pal$^\ast$}
\affiliation{Department of Metallurgical Engineering and Materials Science, Indian Institute of Technology Bombay, Powai, Mumbai 400076, India}
\affiliation{Centre of Excellence in Quantum Information, Computation, Science and Technology, Indian Institute of Technology Bombay, Powai, Mumbai 400076, India} 

\begin{abstract}

The realization of superconducting diodes that operate without external magnetic fields and allow electrical control of polarity is a key goal for the integration of nonreciprocal elements into cryogenic and quantum technologies. Here, we demonstrate a universal and scalable approach to achieving such field-free and electrically reconfigurable Josephson diode functionality. Our method relies on long Josephson junctions with ferromagnetic barriers and asymmetric current injection-—a configuration that inherently breaks both time-reversal and inversion symmetries. We show that the diode polarity is set by an applied bias current and can be reversed using short current pulses, without the need for magnetic fields or thermal cycling. The effect is robust, material-agnostic, and compatible with standard established fabrication processes. Our results provide a practical platform for integrating low-dissipation, programmable diodes into superconducting and quantum electronic circuits.

\end{abstract}

\maketitle

\section{Introduction}
Semiconducting diodes exploit broken inversion symmetry to allow unidirectional charge flow, but their inherent dissipation limits applicability in cryogenic and quantum systems.  With the rise of quantum technologies, there is a growing demand for nondissipative diodes operable at low temperatures. Achieving dissipation less nonreciprocal transport requires breaking both inversion and time-reversal symmetries. This has recently been realized in various ways, both in bulk superconductors \cite{ando2020observation,ilic2022theory,he2022phenomenological,legg2022superconducting,de2023superconducting,hosur2023proximity,daido2022intrinsic,kealhofer2023anomalous,karabassov2022hybrid,lin2022zero,narita2022field,hou2023ubiquitous,satchell2023supercurrent,yun2023magnetic}  and in Josephson junctions (JJs), called Josephson diodes (JDs) \cite{zhang2022general,davydova2022universal,steiner2023diode,lu2023tunable,pal2022josephson,baumgartner2022supercurrent,turini2022josephson,gupta2023gate,ghosh2024high,wu2022field,jeon2022zero,diez2023symmetry,anwar2023spontaneous,golod2022demonstration}.  Initial JD implementations required external magnetic fields \cite{pal2022josephson,baumgartner2022supercurrent,turini2022josephson,gupta2023gate,ghosh2024high}, while more recent approaches have enabled field-free operation by invoking exotic mechanisms—ranging from weak-link polarization \cite{wu2022field}, $\phi_0$-junction physics via magnetic proximity \cite{jeon2022zero}, orbital magnetism in twisted bilayer graphene \cite{diez2023symmetry}, and unconventional superconducting order parameters \cite{anwar2023spontaneous}, to artificially patterned vortices in long junctions \cite{golod2022demonstration}. These mechanisms, while effective, typically rely on a combination of spin-orbit coupling, non-centrosymmetry, proximity-induced magnetism, or nontrivial topology \cite{zhang2022general,davydova2022universal,steiner2023diode,lu2023tunable}. 

Departing from prior approaches that rely on tailored material systems or magnetic control, we propose and demonstrate a universal and scalable method for realizing a Josephson diode--one that operates without an external magnetic field and whose polarity can be reversed electrically.
We show that any long Josephson junction — defined as a junction whose length exceeds the Josephson penetration depth $\lambda_J$ \cite{barone1982physics,tinkham2004introduction} — with a ferromagnetic weak link and asymmetric current injection (achievable experimentally using a cross junction geometry), inherently acts as a Josephson diode. Remarkably, the direction of diode rectification is set entirely by the direction of initial bias current, and can be reversed via additional current pulses. This all-electrical, material-agnostic approach can be implemented using standard Nb-based trilayer junction technology, making it fully compatible with large-scale superconducting circuits and quantum hardware platforms.

\begin{figure*}
	\includegraphics[width=\linewidth]{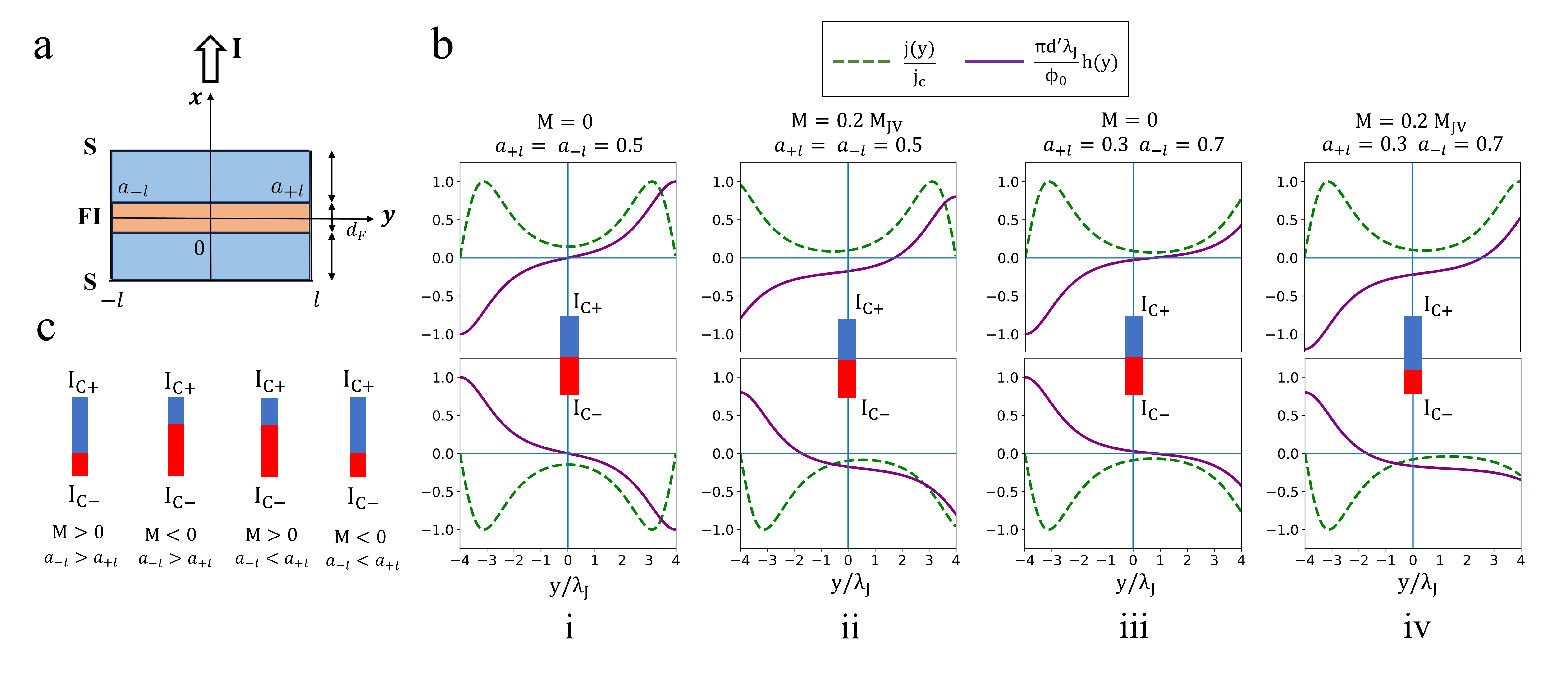}
		\caption{a) Schematic illustration of the JJ.  b) Spatial profiles of local critical current density $j(y)$ (green, dashed) and local magnetic field $h(y)$ (purple, solid) in different scenarios for $L=8\lambda_{J}$ and zero external magnetic field: (i) symmetric JJ with no magnetization  (ii) symmetric JJ with magnetization, (iii) asymmetric JJ with no magnetization, and (iv) asymmetric JJ with magnetization. Nonreciprocity in current, i.e., $I_{c+}\ne I_{c-}$ occurs only in the last scenario. Here, $M$ is normalized by $M_{\text{JV}} = \frac{2 \lambda_{J} j_{c} d'}{d_{F}}$, the value of $M$ for which the junction critical current vanishes, which corresponds to one full Josephson vortex in the JJ. c) Direction of current nonreciprocity switches when reversing magnetization direction ($M\rightarrow -M$) or interchanging the asymmetry parameters ($a_{\pm l} \rightarrow a_{\mp l}$).}
	\label{f1}
\end{figure*}

\section{Theory}
We first establish the basic idea. Consider a JJ formed by sandwiching a ferromagnetic insulator (FI) layer between two s-wave superconducting (S) layers, as shown in Fig. \ref{f1}(a). The layers are stacked in the $x$ direction, with their faces parallel to the $y-z$ plane. The junction has a length $L=2l$ in the $y$ direction. The FI layer is of thickness $d_F$ and is assumed to have a uniform magnetization $M$ along the $z$ direction. A Josephson current $I$ (per unit length in the $z$ direction) flows along the $x$-direction.  To calculate this, we follow the procedure outlined in \cite{owen1967vortex}. The gauge invariant phase difference $\phi(y)$ between the two $S$ layers is related to the local magnetic field $h(y)$ as $\frac{d\phi}{dy} = \frac{2\pi}{\phi_{0}}[h(y)d' + 4\pi Md_{F}]$,
where $d'$   is the width of the junction penetrated by $h(y)$.
Ampere's law relates $h(y)$ with the current density $j(y)$:
$\frac{dh}{dy} = 4\pi j(y)$.
We work in units where $c=\hbar=1$. Combining the two equations with Josephson's current-phase relation $j(y)=j_{c}\text{sin}\phi$, where  $j_{c}$ is the critical current density, we have:
\begin{equation}
    \label{E3}
    \frac{d^{2}\phi}{dy^{2}} = \frac{1}{\lambda_{J}^{2}}\:\text{sin}\phi,
\end{equation}
where $\lambda_{J}=\sqrt{\frac{\phi_{0}}{8\pi^{2} d' j_{c}}}$ is the Josephson penetration depth. Equation \ref{E3} is supplemented with the general boundary conditions \cite{barone1982physics}:
\begin{align}
    \label{E4}
    h(l)+h(-l) &=2H_e -  4\pi (a_{-l}-a_{+l}) I,\\ \label{E5}
    h(l)-h(-l) &= 4\pi I,
\end{align}
with $a_{+l}+a_{-l} = 1$. Here $I=\int_{-l}^{l}j(y)dy$, $H_e$ is the external magnetic field, and the parameters $a_{\pm l}$ characterize the asymmetry in current transport at the edges of the JJ, which are determined by device geometry [\cite{barone1982physics,barone1975current}]--for no asymmetry in current injection, $a_{+l} = a_{-l} = 0.5$. We consider the situation when there is no external magnetic field, i.e., $H_e=0$. To find the critical current $I_c$, one needs to solve Eq.~(\ref{E3}) subject to Eqs.~(\ref{E4}) and (\ref{E5}), and choose the solution that maximizes $I$. Details of the calculation are provided in Appendix A; here we provide our results. Fig. \ref{f1}(b) shows the spatial distribution of $h(y)$ and $j(y)$  corresponding to the critical current, for both positive (+) and negative (-) bias, in various scenarios: (i) a long symmetric nonmagnetic JJ ($a_{+l}=a_{-l}$, $M=0$) (ii) a long symmetric magnetic JJ ($a_{+l}=a_{-l}$, $M\ne 0$), (iii) a long asymmetric nonmagnetic JJ ($a_{+l}\ne a_{-l}$, $M=0$), and (iv) a long asymmetric magnetic JJ ($a_{+l}\ne a_{-l}$, $M\ne 0$). In (i), $j_{+}(y)=-j_{-}(y)$, resulting in an $I_c$ that is the same for both positive and negative bias, i.e., $I_{c+}=I_{c-}$. In (ii) and (iii), the distribution $j(y)$ changes, resulting in a different $I_c$. Nevertheless, in (ii), $j_{+}(y)=-j_{-}(-y)$ and in (iii), $j_{+}(y) = -j_{-}(y)$, both yielding $I_{c+}=I_{c-}$. In (iv) no such symmetry survives, $I_{c+} \neq I_{c-}$, and the system acts as a diode.

\begin{figure*}
    \includegraphics[width=0.9\linewidth]{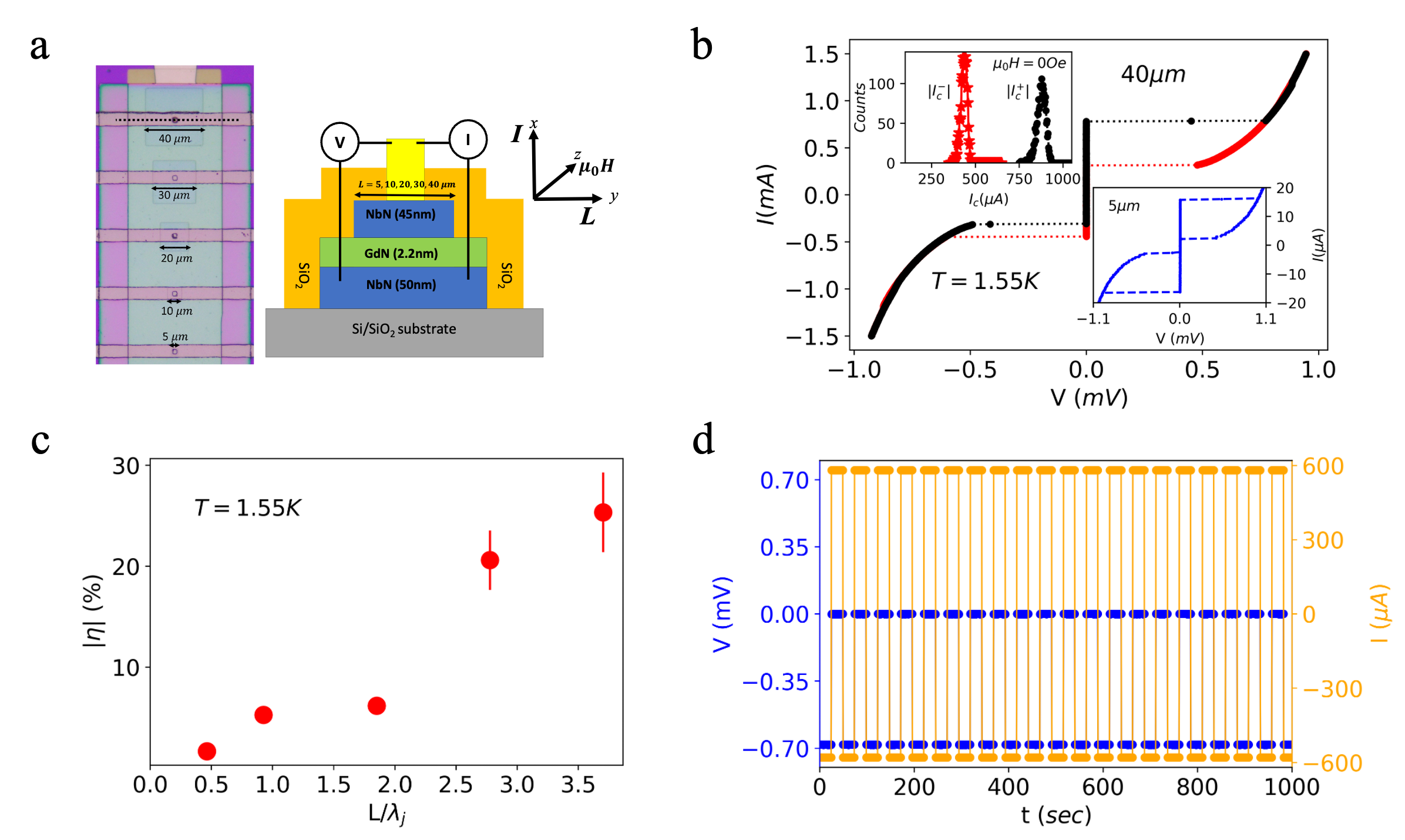}
	\caption{a) Optical micrograph (top view) of the fabricated devices in a single chip, and cartoon cross-sectional diagram of measured junctions with axes definition for field direction b) IV characteristic of $40\:\mu \text{m} \times 40\:\mu \text{m}$ junction, at zero field. Top left inset shows SCD of the same junction for positive and negative critical currents. Bottom Inset shows zero-field IV characteristics of a short $5\:\mu \text{m} \times 5\:\mu \text{m}$ junction at zero field. c) Experimentally obtained efficiency values as a function of junction length. d) Blue line represents half-wave rectified voltage response at zero field, due to an applied square wave current bias signal (orange) to the $40\:\mu \text{m} \times 40\:\mu \text{m}$ long junction.} 
    \label{f2}
\end{figure*}

Insight into the diode effect can be gained by studying the equilibrium configuration of Josephson vortices in the JJ resulting from the self-field $h(y)$, as shown in Fig. \ref{f1}(b). A symmetric nonmagnetic JJ [scenario (i)] features a vortex at $y=l$ and an antivortex at $y=-l$, with equal energetic costs for their formation, and penetrate equally towards the center of the junction. Reversing the current direction exchanges their sides, or equivalently, converts one into the other while preserving their positions. 
Introducing magnetization via the FI barrier [scenario (ii)] or edge asymmetry in current injection [scenario (iii)] breaks this symmetry between the vortex and the antivortex: one becomes energetically favored and penetrates more into the junction. Reversing the current maintains this energetic preference, either causing the vortex and anti-vortex to swap sides [scenario (ii)] or to exchange roles while remaining on the same sides [scenario (iii)]. In scenario (iv), with both magnetization and edge asymmetry present, competition arises between these two tendencies, breaking all symmetries and leading to a nonreciprocal critical current. Importantly, a long junction is essential: In a short junction, $h(y)$ remains nearly constant, preventing vortex / antivortex formation in the first place.

The key requirement, thus, is that any symmetry in the vortex–antivortex configuration, set by the local field profile $h(y)$, must be broken to produce diode-like current asymmetry. Earlier approaches achieved this by externally engineering field inhomogeneity—for example, through spatially varying magnetic fields \cite{krasnov2020josephson} or trapped Abrikosov vortices \cite{golod2022demonstration}. In our case, the required asymmetry arises intrinsically from the combined action of barrier magnetization and asymmetric current injection in a long Josephson junction, with the former breaking time reversal symmetry and the latter breaking inversion symmetry \cite{zinkl2022symmetry}.

A corollary of the above discussion is that the current non-reciprocity direction depends on the magnetization direction and $a_{+l}/a_{-l}$. Switching either reverses the non-reciprocity direction, whereas simultaneously changing both leaves it intact, as shown schematically in Fig.~\ref{f1}(c).

\section{Experimental results and discussion}
We now present an experimental realization of the idea espoused above using NbN/GdN/NbN spin-filter \cite{senapati2011spin,pal2014pure} (see comment \footnote{Such a JJ is known to have a pure second-harmonic current-phase relationship\cite{pal2014pure}: $j(y)=j_c\mathrm{sin}(2\phi)$. However, a rescaling of $\phi\rightarrow\phi/2$, $y\rightarrow y/\sqrt{2}$, and $I\rightarrow I/\sqrt{2}$ leaves Eq.~\ref{E3} invariant. Consequently, our theoretical analysis remains applicable to such a system.}) mesa-type long JJs in cross geometry (which intrinsically allows for asymmetric current injection\cite{barone1975current,owen1967vortex,kuplevakhsky2007exact,krasnov1997fluxon}). The junctions are fabricated using a standard multi-step lithography process\cite{blamire2012spin,huggins1985preparation}. An optical micrograph of the central device area with varying size mesas on top of an NbN ground line, wired with Nb cross bridges, and a cartoon cross-section of a typical junction are shown in Fig.~\ref{f2}(a). In Fig.~\ref{f2}(b) we show the zero-field IV curve of the  $40\mu m$ long junction along with a switching current distribution (SCD) histogram recorded at zero field, as the top left inset; where a nonreciprocal effect is visible. The bottom right inset shows the zero-field cooled IV of the short $5\mu m$ junction, where the positive and negative critical currents are almost identical.  In Fig. \ref{f2}(c) we present the experimentally obtained dependence of $\eta$ on $L$. As expected, $\eta$ increases as $L/\lambda_J$ grows. [See Appendix A and Appendix D for calculation of $\lambda_J$, Appendix C for representative IV curves of all other junctions, and Appendix B for Fraunhofer patterns of long and short junctions]. In Fig.~\ref{f2}(d), we demonstrate half-wave rectification of a square wave current signal with the $40\mu m$ long  JJ. This demonstrates that our long magnetic JJ is indeed a zero-field JD as proposed.  

We next turn to the question of whether the diode polarity can be switched. As discussed earlier, reversing the direction of the barrier’s net magnetization would reverse the direction of nonreciprocity, thereby switching the diode polarity. While this could in principle be achieved by applying an external magnetic field, it is unnecessary—the barrier can also be polarized solely by the bias current via the magnetic field it generates. This eliminates the need for any external field and provides a direct electrical knob to control the diode polarity.

\begin{figure}
\includegraphics[width=\linewidth]{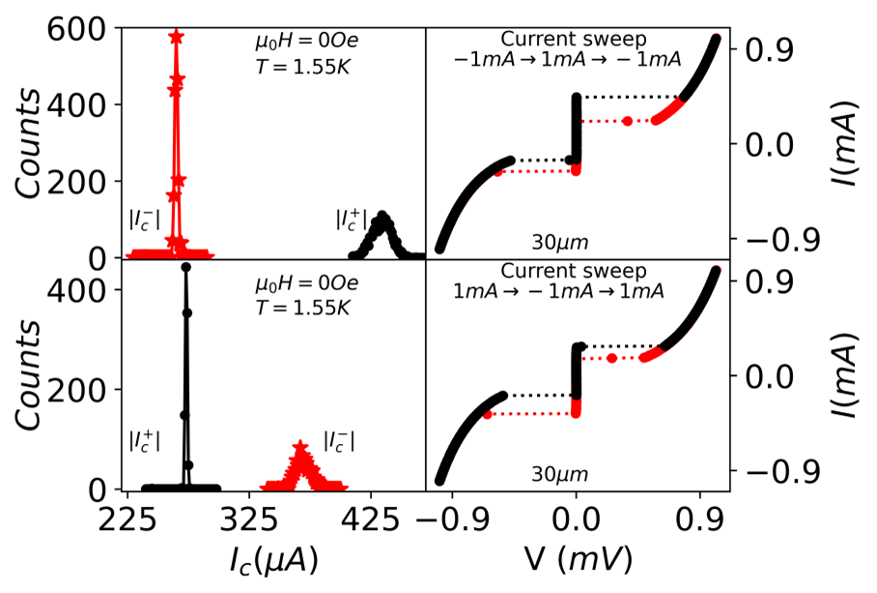}
	\caption{Demonstration of diode polarity reversal through IV and SCD characteristics of the $30\:\mu \text{m} \times 30\:\mu \text{m}$ junction, at zero field for two different current sequences: upper right presents IV for current sequence $-1\: \text{mA} \rightarrow 1\: \text{mA}\rightarrow -1\: \text{mA}$, followed by corresponding SCD shown in upper left (2000 points). Lower right presents IV for current sequence $1\: \text{mA} \rightarrow -1\: \text{mA}\rightarrow 1\: \text{mA}$, followed by corresponding SCD shown in lower left (1000 points). Junction was heated above ordering temperature of GdN and cooled down between the two IV and SCD measurements.}  
	\label{f3}
\end{figure}

We first demonstrate this in the $30\mu m$ junction (Fig.~\ref{f3}), where the zero-field IV of the junction is recorded after two separate cooling cycles. In the first cycle, we record the current-biased IV of the junction by starting at -1mA (upper panels of Fig. \ref{f3}). After recording an IV, we measured the SCD for the junction (upper left panel of Fig. \ref{f3}). The junction shows higher magnitude of positive critical current in this case. In order to remove any magnetic history of the GdN barrier, we heat our junction above the GdN ordering temperature (35K), to 50K and cool it down again to 1.55K and measure the current-biased IV of the junction by starting at +1mA and follow it up with an SCD measurement (bottom panels of Fig.~\ref{f3}). It is evident from these measurements that the negative critical current magnitude now exceeds the positive critical current, as anticipated. This confirms the picture presented above and, in turn,  demonstrates that our JD is polarity reversible, in addition to operating at zero field.
 
\begin{figure}
	\includegraphics[width=0.9\linewidth]{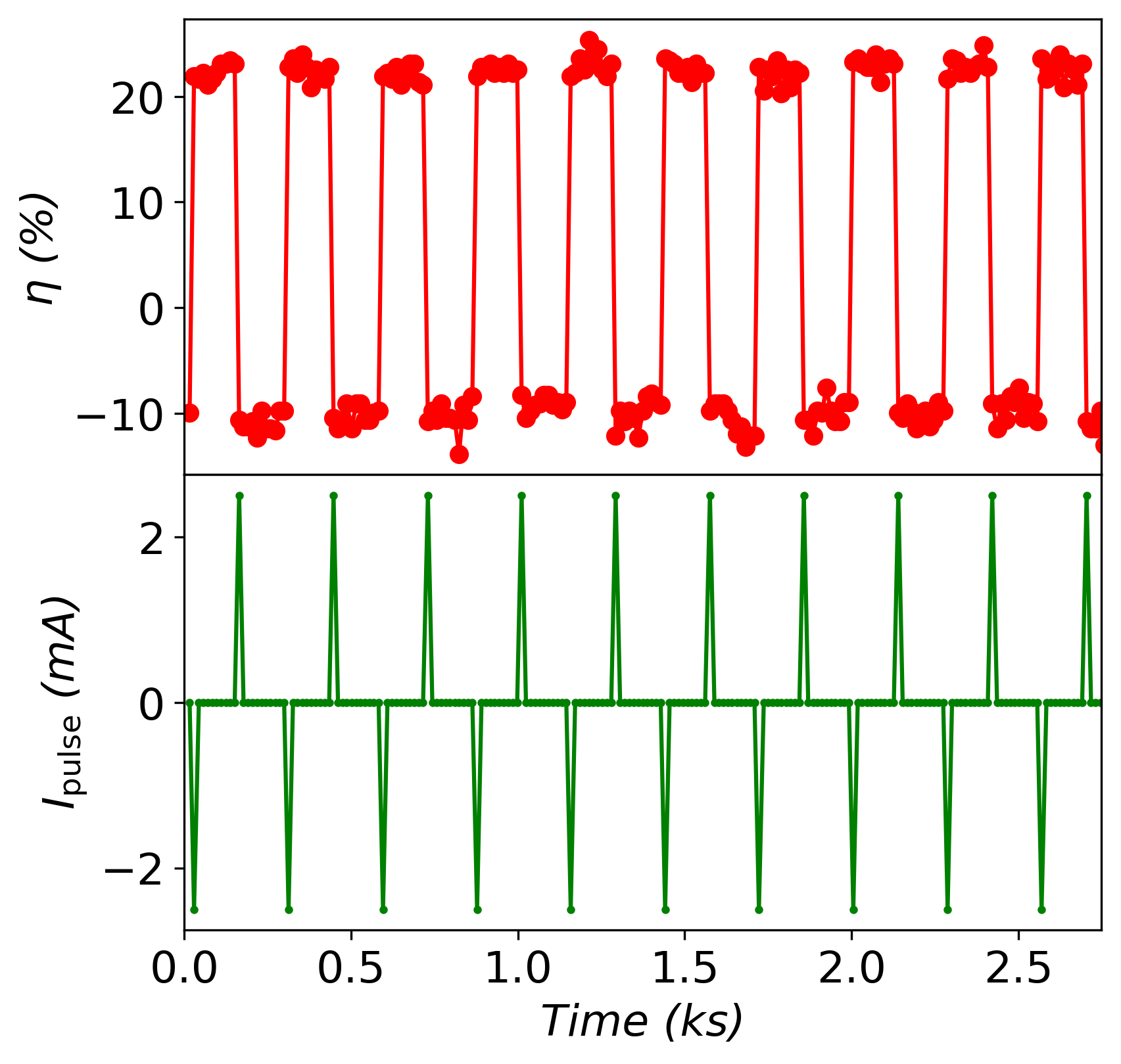}
	\caption{Demonstration of diode polarity reversal using 100ms current pulses to the  junction. Upper panel shows variation of diode efficiency ($\eta$)  after application of each applied current pulse. IVs were measured continuously after each pulse, and diode efficiency was extracted for each IV measurement. Bottom panel shows the sequence of current pulses that were applied to the junction as a function of time.}
	\label{f4}
\end{figure}

Building on the above, we next demonstrate that the diode polarity can be reversed purely by applying current pulses, without resetting the device’s magnetic history via thermal cycling. Indeed, the induced magnetization of the barrier can be reversed by applying a current pulse of sufficient magnitude in the opposite direction, thereby switching the diode polarity. We demonstrate this experimentally in the same junction by performing polarity reversals with current pulses of amplitude $\pm$2.5mA. The results are shown in Fig.~\ref{f4}. After each current pulse application, the JD undergoes a sharp polarity reversal and is able to hold its changed state remarkably well. This demonstrates that the polarity can be programmed on demand using only electrical pulses.

The reproducibility, persistence, and binary nature of these reversals are consistent with controlled switching of the junction’s internal magnetic state. However, the microscopic mechanism underlying these current-pulse-driven magnetization reversals likely involves a complex interplay of magnetic domains, proximity effects, and related processes, and developing a complete theoretical understanding of this mechanism remains an important direction for future work.

To the best of our knowledge, only a few prior works have demonstrated polarity reversibility in Josephson diodes: one by altering the magnetic field history of a $\phi_0$ junction \cite{jeon2022zero}, and another by trapping or expelling a vortex in a planar junction structure \cite{golod2022demonstration}. The former requires an external magnetic field and is therefore not field-free. The latter achieves field-free operation but relies on a complex multi-terminal geometry involving two Josephson junctions and a deliberately fabricated vortex trap between them. In contrast, our approach leverages the long-junction regime and integrates ferromagnetic magnetization directly within the mesa-type junction barrier. These two features allow for a significantly simplified device architecture—one that can be readily implemented using standard Nb-based superconducting foundry processes. As a result, our device is distinguished by its fabrication simplicity, scalability, and straightforward, field-free operation without the need for additional terminals.

\section{Conclusion}
In conclusion, we have demonstrated a novel and robust route to achieving a zero-field Josephson diode with electrically switchable polarity. Our approach is based on a long magnetic Josephson junction with a geometry that enables asymmetric current injection. This mechanism is universal, relying only on minimal and broadly realizable conditions. It is independent of the microscopic details of the superconducting or magnetic materials and leverages asymmetries that naturally arise in standard cross junction layouts. Importantly, device fabrication is fully compatible with established industrial processes such as sputtering and optical lithography, making this method readily scalable for integration into superconducting electronics and quantum technologies.

\begin{acknowledgments}

The work was financially supported by a grant from National Supercomputing Mission, India under Grant Number - DST/NSM/R$\&$D Exascale/2021/10. HKP would like to thank DST SERB, India for financial support via Grant No. CRG/2021/005453.
\\\\
\noindent $^\dagger$These authors contributed equally to this work.
\\
\noindent$^\ast$avradeep@iitb.ac.in

\end{acknowledgments}

\begin{widetext}
\begin{appendix}
\section{Theory of diode effect in a ferromagnetic long Josephson junction with asymmetric injection of current}
\label{SM1}
Our main proposal to achieve a diode effect, as stated in the main text, is to use a ferromagnetic long Josephson junction with an asymmetric injection of current. In this section, we substantiate this claim through explicit calculations. The classic problem of the variation of current in Josephson junctions subjected to a homogeneous magnetic field has been studied extensively, both numerically and analytically. Here, we extend it to the case where the barrier is a single-domain ferromagnetic insulator of uniform magnetization. We consider the setup presented in Fig. \ref{F1}.

Two identical semi-infinite superconductors with width $L=2l$ are placed with their faces parallel to the $y-z$ plane, at a separation $d_{F}$ from each other. A ferromagnetic material of thickness $d_{F}$ and length $L$ is sandwiched between the faces of the superconductors. The $x$-axis lies perpendicular to the layering direction. A constant external homogeneous magnetic field $H_{e}$ is applied along the $z$ direction and the transport current $I$ (per unit length in the $z$ direction) is injected along the $x$ axis. \\

The dynamics of the system is specified by the gauge invariant phase difference $\phi(y)$ across the junction. The magnetic field at the junction has three contributions:
\begin{itemize}
    \item[i)] The external field $H_{e}$
    \item[ii)] The induced field due to the Josephson current
    \item[iii)] Magnetization of the ferromagnet 
\end{itemize}

Let $h(y)$ denote the resultant magnetic field due to (i) and (ii). $h(y)$ penetrates the junction in a region of width $d'$ given by:
\begin{equation}
    \label{S1}
    d'  = \mu_{r}d_{F} + \lambda_{\text{bot}}\:\text{coth}\left(\frac{d_{\text{bot}}}{\lambda_{\text{bot}}} \right) + \lambda_{\text{top}}\:\text{coth}\left(\frac{d_{\text{top}}}{\lambda_{\text{top}}} \right),
\end{equation}
$\lambda_{\text{top}} (\lambda_{\text{bot}})$ being being the London penetration depth of the top (bottom) superconductor. Consider an elemental rectangular region of dimensions $d' \Delta y$, at a distance $y$ from the origin. The magnetic flux through the elemental region is (in Gaussian units)
\begin{equation} 
    \label{S2}
    \Delta \varphi = h(y)d'\Delta y + 4\pi M d_{F}\Delta y,
\end{equation}
where $M$ is the magnetization. Note that $M$, in general, is a function of $h(y)$. For simplicity, in what follows, we assume that this is a constant (this suffices to understand the qualitative features of the experiment). The gauge invariant phase difference $\phi(y)$ across the junction is related to the flux $\Delta \varphi$ as \cite{tinkham2004introduction}:
\begin{equation}
    \label{S3}
    \Delta \phi = \frac{2\pi}{\phi_{0}}[h(y)d'\Delta y + 4\pi M d_{F}\Delta y],
\end{equation}
where $\phi_{0}=h/2e$ is the superconducting flux quantum. In the limit $\Delta y \rightarrow 0$:
\begin{equation}
\label{S4}
\frac{d \phi}{d y} = \frac{2\pi}{\phi_{0}}[h(y)d'+4\pi M d_{F}].
\end{equation}
Maxwell's equations relate the field $h(y)$ with the free current density $j(y)$ as
\begin{equation}
\label{S5}
    \frac{dh}{dy} = \frac{4\pi}{c}j(y).
\end{equation}
The current density $j(y)$ is related to the phase difference $\phi(y)$, following Josephson's current phase relation:
\begin{equation}
\label{S6}
    j(y) = j_{c}\:\text{sin}\phi.
\end{equation}
Differentiating Eq.~$\ref{S4}$ and substituting  Eqs.~\ref{S5}, \ref{S6} lead to the static Sine-Gordon equation:
\begin{equation}
\label{S7}
    \frac{d^{2} \phi}{dy^{2}} = \frac{1}{\lambda_{J}^{2}}\text{sin}\phi,
\end{equation}
where  $\lambda_{J} = \sqrt{\frac{\phi_{0}c}{8\pi^{2}j_{c}d'}}$ is the Josephson penetration depth. Henceforth, we rescale $y\rightarrow y/\lambda_J$ and use $\hbar=c=1$. Then, Eq.~\ref{S7} becomes 
\begin{equation}
\frac{d^{2} \phi}{dy^{2}} = \text{sin}\phi .
\label{normsingor}
\end{equation}
\begin{figure}
	\includegraphics[width=0.5\linewidth]{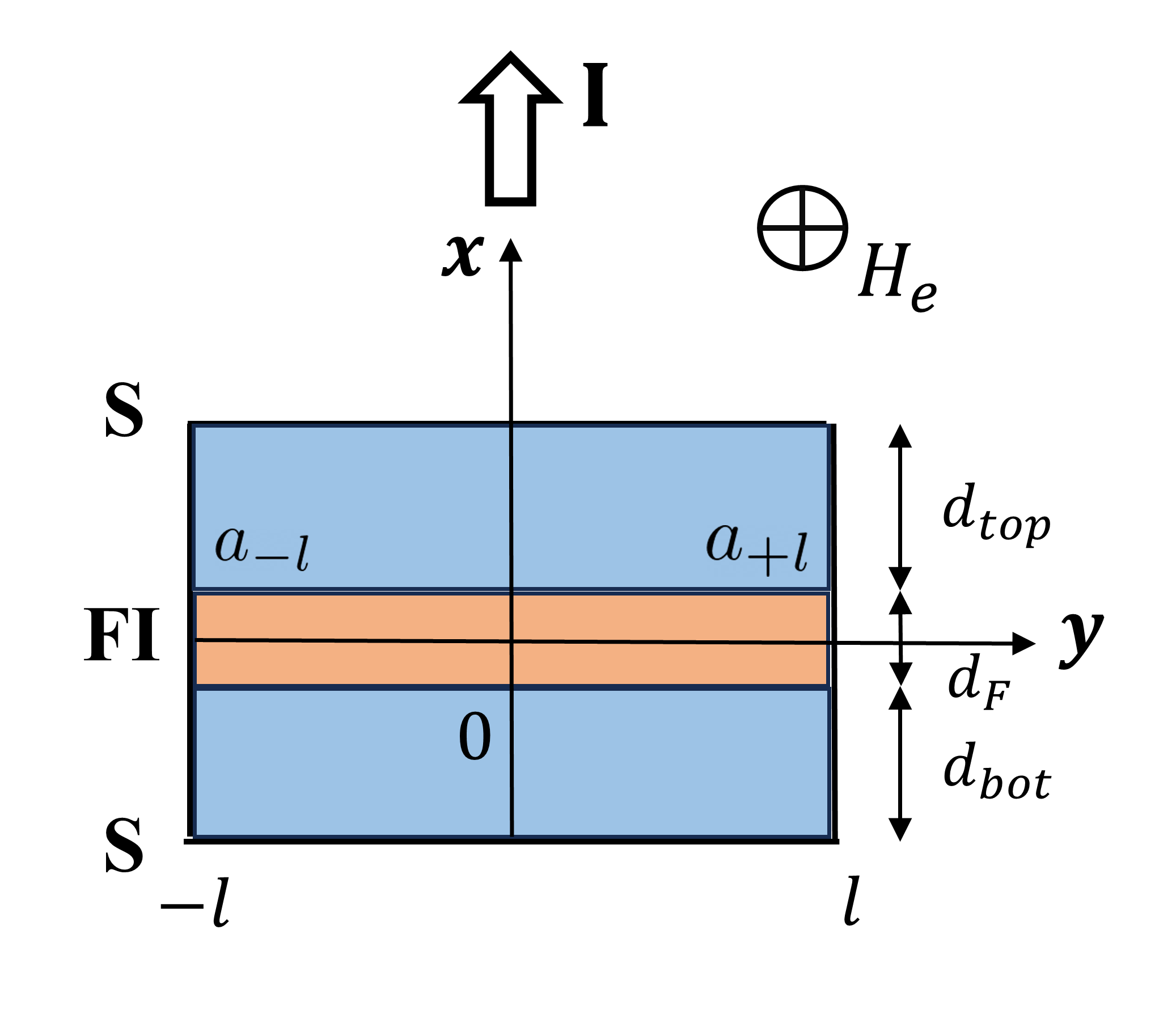}
		\caption{Schematic figure of a Josephson junction considered in this work: $d_{F}$ is the thickness of the ferromagnetic material. A homogeneous magnetic field $H_{e}$ is directed into the plane of paper. The transport supercurrent $I$ is along the $x$ axis.} 
	\label{F1}
\end{figure}
To find the solution to the above equation, we need boundary conditions. To that effect, we use Eq.~\ref{S4} at the endpoints:
\begin{align}
    \label{S9}
    \frac{d \phi}{d y}\bigg{|}_{l} &= \frac{2\pi}{\phi_{0}}\lambda_{J}[d'h(l)+4\pi Md_{F}], \\
    \label{S10}
     \frac{d \phi}{d y}\bigg{|}_{-l} &= \frac{2\pi}{\phi_{0}}\lambda_{J}[d'h(-l)+4\pi Md_{F}].
\end{align}
Integrating Eq.~\ref{S5} along $y$, along with the continuity of the magnetic field at the junction edges yield the boundary conditions for the induced field $h(y)$, which can be written as \cite{owen1967vortex}:
\begin{align}
    \label{S11}
    h(l)+h(-l) =  2H_{e}, \\ \label{S12}
    h(l) - h(-l) = 4\pi I ,
\end{align} 
where $I=\int_{-L}^{L}dy\:j(y)$.
Substitution of Eqs.~\ref{S9} and \ref{S10} into Eqs.~\ref{S11} and \ref{S12} leads to the expressions for $H_{e}$ and $I$ in terms of the phase derivatives at the boundary:
\begin{align}
   \label{S13}
    H_{e} &= \frac{\phi_{0}}{4\pi \lambda_{J}d'}\left[\frac{d\phi}{dy}\bigg{|}_{l} + \;\frac{d\phi}{dy}\bigg{|}_{-l} \right] -H_{0},\;\;\;\;\;\;\;H_{0} = \frac{4\pi Md_{F}}{d'}, \\ \label{S14}
    I &= \frac{\phi_{0}}{8\pi^{2}\lambda_{J}d'}\left[\frac{d\phi}{dy}\bigg{|}_{l} - \;\frac{d\phi}{dy}\bigg{|}_{-l} \right].
\end{align}
So far, we have assumed that the junction is symmetric. However, in the experiment, the junction has a cross-type geometry. In this case, it is well known that there is an asymmetric current splitting at the junction edges and, as a result, the boundary conditions are modified as  \cite{barone1975current,vaglio1976approximate},
\begin{align}
    H_{e}&=\frac{\phi_{0}}{2\pi \lambda_{J}d'}\left[a_{-l}\frac{d\phi}{dy}\bigg{|}_{l} + a_{-l}   \frac{d\phi}{dy}\bigg{|}_{+l}\right] -H_{0}, \label{S15}\\
     I &= \frac{\phi_{0}}{8\pi^{2}\lambda_{J}d'}\left[\frac{d\phi}{dy}\bigg{|}_{l} - \;\frac{d\phi}{dy}\bigg{|}_{-l} \right], \label{S16}\\ 
     & a_{\pm l} \geq 0,\;\;\;a_{+l}+a_{-l}=1\;\;\;.
\end{align}
The parameters $a_\pm$ measure the degree of the asymmetry in current injection at the two edges.
We note, however, that the above boundary conditions are ill-posed since they are not sufficient for the uniqueness of a solution to the Sine-Gordon equation \ref{normsingor} \cite{courant1962methods}. As suggested in Refs.~\onlinecite{kuplevakhsky2004topological,kuplevakhsky2006static,kuplevakhsky2007exact} among all the solutions, only those that minimize a certain Gibbs free-energy functional are physically realizable. In what follows, we have also adopted the same strategy. Further details may be found in the above references. Thus, the problem of finding the current as a function of the external field reduces to solving Eq.~\ref{normsingor} subject to the conditions \ref{S15} and \ref{S16}.
To that effect, we multiply both sides of Eq.~\ref{normsingor} by $\frac{d\phi}{dy}$ and integrate to obtain,
\begin{equation}
    \label{S18}
    \frac{1}{2}\left[ \frac{d\phi}{dy}\right]^{2}+cos\:\phi = C,\;\;\;\;\;\;\; -1\leq C < \infty ,
\end{equation}
C being the constant of integration. The solutions are divided into two types, depending on the domain of $C$, and are expressed in terms of Jacobi elliptic functions:\\

    \textbf{Type I}: $\mathbf{-1 \leq C \leq 1.}$ 
    \begin{align*}
        & \phi_{\pm}(y) = (2n+1)\pi \pm 2 sin^{-1}[k\:sn(y-y_{0},k)] ,\\
        & n \in \mathbb{Z},\;\;\; -K(k) \leq y_{0} < K(k),\;\;\; k^{2} \equiv \frac{1+C}{2},\;\;\;0\leq k \leq 1 .
    \end{align*}
    \textbf{Type II}: $\mathbf{1 \leq C < \infty}$.
    \begin{align*}
        & \phi_{\pm}(y) = (2n+1)\pi \pm 2\:am\left(\frac{y-y_{0}}{k},k \right), \\
        & -kK(k) \leq y_{0} < kK(k),\;\;\; k^{2} \equiv \frac{2}{1+C},\;\;\;0<k\leq1 .
    \end{align*}
    
The functions $am\:u$ and $sn\:u$ are the Jacobi elliptic amplitude and elliptic sine respectively and $K(k)$ is the complete elliptic integral of the first kind:
    \begin{equation*}
        K(k)=\int_{0}^{\pi/2}d\theta \: \frac{1}{\sqrt{1-k^{2}sin^{2}\theta}}.
    \end{equation*}
The signs of $\phi(y)$ ($\phi_{+} \text{ or } \phi_{-}$), and the value of the constants $k$ and $y_{0}$ are determined from the boundary conditions \ref{S15} and \ref{S16}. For simplicity, let us restrict our analysis to the first quadrant ($I \geq 0,\; H_{e}+H_{0} \geq 0$). The solutions can readily be generalized to the other quadrants. \\ \\
 
\textbf{Solutions of Type I.}
The physically stable solutions are:
\begin{equation}
\label{S19}
    \phi_{s}(y) = 2\:cos^{-1}\left[k\frac{cn(y+\beta,k)}{dn(y+\beta,k)} \right],\;\;\; k \in [k_{c},1),\;\;\;\beta \in [0,\beta_{c}],
\end{equation}
Here $cn$ and $dn$ are the Jacobi elliptic cosine and delta amplitudes, respectively. The above-mentioned values for $k$ and $\beta$ denote their domains of stability. The boundary of the stability region for $k_{c}=k_{c}(L)$ is determined by the relation \cite{kuplevakhsky2007exact}:
\begin{equation}
    \label{S20}
    cn(l,k_{c})[-E(l,k_{c})+(1-k_{c}^{2})l]+(1-k_{c}^{2})sn(l,k_{c})dn(l,k_{c}) = 0.
\end{equation}
Here $E(\psi,k)$ is the incomplete elliptic integral of the first kind with amplitude $\psi$ and modulus $k$. The stability boundary $\beta_{c} = \beta_{c}(k)$ is given by the solution to the following functional equation \cite{kuplevakhsky2007exact}:
\begin{align*}
    &cn(l+\beta_{c},k)cn(l-\beta_{c},k)[-E(l+\beta_{c},k)-E(l-\beta_{c},k)+(1-k^{2})l]\\ 
    &+(1-k^{2})[sn(l+\beta_{c},k)cn(l-\beta_{c},k)dn(l+\beta_{c},k) \\
    &+sn(l-\beta_{c},k)cn(l+\beta_{c},k)dn(l-\beta_{c},k)] = 0,\;\;\;\;\;k \in [k_{c},1).
\end{align*}
The local field and current densities are obtained from equations \ref{S4} and \ref{S6} respectively:
\begin{align*}
    h_{s}(y) &= \frac{\phi_{0}}{\pi d'\lambda_{J}}\frac{k}{\sqrt{1-k^{2}\frac{cn^{2}(y+\beta,k)}{dn^{2}(y+\beta,k)}}}sn(y+\beta,k)\left[1-k^{2}\frac{cn^{2}(y+\beta,k)}{dn^{2}(y+\beta,k)} \right] - \frac{4\pi M d_{F}}{d'}, \\
    j_{s}(y) &= 2j_{c}k \frac{cn(y+\beta,k)}{dn(y+\beta,k)}\sqrt{1-k^{2}\frac{cn^{2}(y+\beta,k)}{dn^{2}(y+\beta,k)}}.
\end{align*}
Substituting equation \ref{S19} in the boundary conditions \ref{S15}, \ref{S16} yields:
\begin{align}
    H_{e}+H_{0} &= \frac{\phi_{0}}{\pi \lambda_{J}d'} k\sqrt{1-k^{2}}\left[a_{-l}\frac{sn(l+\beta,k)}{dn(l+\beta,k)} - a_{+l}\frac{sn(l-\beta,k)}{dn(l-\beta,k)} \right] \label{S21}\\
    I &= \frac{\phi_{0}}{4\pi^{2}\lambda_{J}d'} k\sqrt{1-k^{2}}\left[\frac{sn(l+\beta,k)}{dn(l+\beta,k)} +\frac{sn(l-\beta,k)}{dn(l-\beta,k)} \right] \label{S22}\\
    & k \in [k_{c},1), \;\;\; \beta \in [0,\beta_{c}]
\end{align}
Note that $I$ corresponds to the critical current $I_{c}$ if we set $\beta = \beta_{c}$ in the above equations. This gives the desired dependence of $I_{c}$ with $H_{e}$.\\

\textbf{Solutions of Type II.}
This class of solutions is further divided into two types, depending on the stability domains determined by $k$ and $\alpha$:
\begin{align}
    \phi_{pe}(y) &= (p-1)\pi + 2\:am\left(\frac{y}{k}+K(k)+\alpha,k \right),\;\;\; \alpha \in [0,\alpha_{c}],\;\;\;p = 2m\;\;(m=0,1,...),\label{phipey} \\ 
    \phi_{po}(y) &= p\pi + 2\:am\left(\frac{y}{k}+\alpha,k \right),\;\;\;\alpha \in [0,\alpha_{c}],\;\;\;p=2m+1\;\;(m=0,1,...).
    \label{phipoy}
\end{align}
The values of $k$ for which the respective solutions are stable are indexed by the integer $p$ as:
\begin{align}
    p&=0,\;\;\;k\in(k_{1},1),\\
    p&=1,2,...,\;\;\;k\in(k_{p+1},k_{p}].
\end{align}
The points $k_{p},\:p=1,2,.....$ are the roots of the equations:
\begin{equation}
    \label{S28}
    pk_{p}K(k_{p}) = l,\;\;\;p=1,2,...
\end{equation}
The boundaries of the stability regions $\alpha_{c}=\alpha_{c}(k)$ for equation \ref{phipey} are given by \cite{kuplevakhsky2007exact}
\begingroup \makeatletter\def\f@size{10}\check@mathfonts
\def\maketag@@@#1{\hbox{\m@th\small\normalfont#1}}%
\begin{align}
    & k^{2}sn\left(\frac{l}{k}+\alpha_{c},k\right)sn\left(\frac{l}{k}-\alpha_{c},k\right)cn\left(\frac{l}{k}+\alpha_{c},k\right)cn\left(\frac{l}{k}-\alpha_{c},k\right)\left[E\left(\frac{l}{k}+\alpha_{c},k\right)+E\left(\frac{l}{k}-\alpha_{c},k\right) \right]\\
    &+sn\left(\frac{l}{k}-\alpha_{c},k\right)cn\left(\frac{l}{k}-\alpha_{c},k\right)dn^{3}\left(\frac{l}{k}+\alpha_{c},k\right)+sn\left(\frac{l}{k}+\alpha_{c},k\right)cn\left(\frac{l}{k}+\alpha_{c},k\right)dn^{3}\left(\frac{l}{k}-\alpha_{c},k\right) = 0,
\end{align}
\endgroup
and for equation \ref{phipoy} are given by \cite{kuplevakhsky2007exact}
\begingroup \makeatletter\def\f@size{10}\check@mathfonts
\def\maketag@@@#1{\hbox{\m@th\small\normalfont#1}}
\begin{align*}
    & \frac{k^{2}}{1-k^{2}}sn\left(\frac{l}{k}+\alpha_{c},k\right)sn\left(\frac{l}{k}-\alpha_{c},k\right)cn\left(\frac{l}{k}+\alpha_{c},k\right)cn\left(\frac{l}{k}-\alpha_{c},k\right) \\
    &\times  \left[E\left(\frac{l}{k}+\alpha_{c},k\right)+E\left(\frac{l}{k}-\alpha_{c},k\right)
      -k^{2}\left[\frac{sn\left(\frac{l}{k}+\alpha_{c},k\right)cn\left(\frac{l}{k}+\alpha_{c},k\right)}{dn\left(\frac{l}{k}+\alpha_{c},k\right)} + \frac{sn\left(\frac{l}{k}-\alpha_{c},k\right)cn\left(\frac{l}{k}-\alpha_{c},k\right)}{dn\left(\frac{l}{k}-\alpha_{c},k\right)} \right] \right] \\
      & - \frac{sn\left(\frac{l}{k}-\alpha_{c},k\right)cn\left(\frac{l}{k}-\alpha_{c},k\right)}{dn\left(\frac{l}{k}+\alpha_{c},k\right)} - \frac{sn\left(\frac{l}{k}+\alpha_{c},k\right)cn\left(\frac{l}{k}+\alpha_{c},k\right)}{dn\left(\frac{l}{k}-\alpha_{c},k\right)} = 0.
\end{align*}
\endgroup
The local fields and current densities for the even and odd solutions are given as:
\begin{align}
    h_{pe}(y) &= \frac{\phi_{0}}{\pi d' \lambda_{J}}\frac{1}{k} dn\left(\frac{y}{k}+K(k)+\alpha_{c},k \right) - \frac{4\pi M d_{F}}{d'}, \\
    h_{po}(y) &= \frac{\phi_{0}}{\pi d' \lambda_{J}}\frac{1}{k} dn\left(\frac{y}{k}+\alpha_{c},k \right) - \frac{4 \pi M d_{F}}{d'}, \\
    j_{pe}(y) &= -2j_{c}\:sn\left(\frac{y}{k}+K(k)+\alpha_{c},k \right)cn\left(\frac{y}{k}+K(k)+\alpha_{c},k \right),\\
    j_{po}(y) &= -2j_{c}\:sn\left(\frac{y}{k}+\alpha_{c},k \right)cn\left(\frac{y}{k}+\alpha_{c},k \right).
\end{align}
Substituting Eq.~\ref{phipey} in \ref{S15} and \ref{S16}, we have
\begin{align}
    H_{e}+H_{0} &= \frac{\phi_{0}}{\pi \lambda_{J}d'}\frac{\sqrt{1-k^{2}}}{k}\left[a_{-l}dn^{-1}\left(\frac{l}{k}+\alpha_{c},k\right)+a_{+l}dn^{-1}\left(\frac{l}{k}-\alpha_{c},k\right) \right] \label{S35}\\
    I &= \frac{\phi_{0}}{4\pi^{2}\lambda_{J}d'}\frac{\sqrt{1-k^{2}}}{k}\left[dn^{-1}\left(\frac{l}{k}+\alpha_{c},k\right)-dn^{-1}\left(\frac{l}{k}-\alpha_{c},k\right) \right] \label{S36}\\
    \alpha &\in [0,\alpha_{c}],\;\;\;p=2m,\;\;m=(0,1,...).
\end{align}
Similarly, substituting Eq.~\ref{phipoy} in \ref{S15} and \ref{S16}, we have:
\begin{align}
     H_{e}+H_{0} &= \frac{\phi_{0}}{\pi \lambda_{J}d'}\frac{1}{k}\left[a_{-l}dn\left(\frac{l}{k}+\alpha_{c},k\right)+a_{+l}dn\left(\frac{l}{k}-\alpha_{c},k\right) \right] \label{S38}\\
    I &= \frac{\phi_{0}}{4\pi^{2}\lambda_{J}d'}\frac{1}{k}\left[dn\left(\frac{l}{k}+\alpha_{c},k\right)-dn\left(\frac{l}{k}-\alpha_{c},k\right) \right] \label{S39}\\
    \alpha &\in [0,\alpha_{c}],\;\;\;p=2m+1,\;\;m=(0,1,...)
\end{align}
The above sets of expressions give a complete set of physically relevant solutions for $H_e$, $I$, $j(y)$, and $h(y)$ in the first quadrant ($H_{e}+H_{0}\geq 0, I \geq 0$). The results can be readily extended to the other quadrants. The overall sign of the phase (choosing $\phi_{+}$ or $\phi_{-}$) is determined from the boundary conditions, and hence depends on which quadrant the values of $H_{e}+H_{0}$ and $I$ reside. Further, note that changing $\alpha \rightarrow -\alpha$ or $\beta \rightarrow -\beta$ doesn't affect the stability conditions for both $\phi_{s}$ and $\phi_{p}$. For a fixed $|H_{e}|$ and $|I|$, the solutions in the other quadrants are obtained through the use of the following symmetry relations:
\vspace{0.2 cm}

\begin{itemize}
    \item[1)] $\mathbf{H_{e}+H_{0} \leq 0.\;I \geq 0}$:
    \begin{equation}
    \label{S41}
        \phi_{s},\beta \rightarrow \phi_{s},-\beta,\;\;\;\;\;\;\phi_{p},\alpha \rightarrow -\phi_{p},-\alpha
    \end{equation}
    \item[2)] $\mathbf{H_{e}+H_{0} \geq 0.\;I \leq 0}$:
    \begin{equation}
    \label{S42}
        \phi_{s},\beta \rightarrow -\phi_{s},-\beta,\;\;\;\;\;\;\phi_{p},\alpha \rightarrow \phi_{p},-\alpha
    \end{equation}
    \item[3)] $\mathbf{H_{e}+H_{0} \leq 0.\;I \leq 0}$:
    \begin{equation}
        \label{S43}
        \phi_{s},\beta \rightarrow -\phi_{s},\beta,\;\;\;\;\;\;\phi_{p},\alpha \rightarrow -\phi_{p},\alpha
    \end{equation} 
\end{itemize} 
\vspace{0.1 cm}

We now plot the variation of critical current $I_{c}$ vs the external field $H_{e}$ for various cases. \\ \\
\textbf{Case 1: Fig.~\ref{F2}}(a). A short junction (l=0.2) with symmetric current injection   ($a_{+l}=a_{-l}=0.5$) and a non-magnetic barrier ($M=0$).\\
\textbf{Case 2: Fig.~\ref{F2}}(b). A long junction (l=2) with symmetric current injection      ($a_{+l}=a_{-l}=0.5$) and a non-magnetic barrier ($M=0$).\\
\textbf{Case 3: Fig.~\ref{F2}}(c). A long junction (l=2) with symmetric current injection    ($a_{+l}=a_{-l}=0.5$) and a magnetic barrier ($M\ne 0$).\\
\textbf{Case 4: Fig.~\ref{F2}}(d). A long junction (l=2) with asymmetric current injection ($a_{+l}\ne a_{-l}$) and a magnetic barrier ($M\ne 0$).\\ \\

\begin{figure}  
\includegraphics[width=0.99\linewidth]{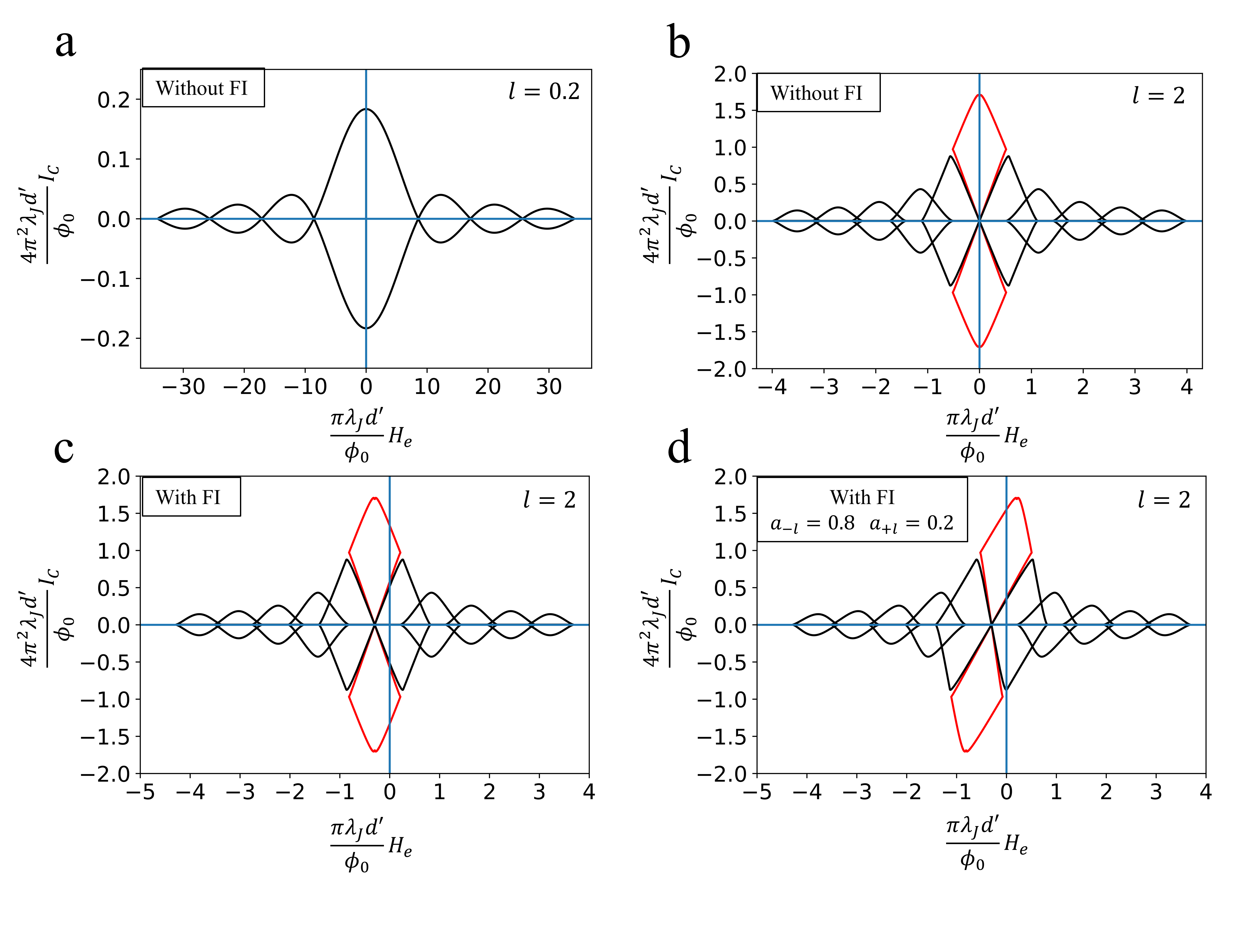}
\caption{Variation of the critical current $I_{c}$ with the external magnetic field $H_{e}$ according to Eqs.~\ref{S21}, ~\ref{S22}, ~\ref{S35}, ~\ref{S36}, ~\ref{S38} and ~\ref{S39} for a) A short junction (l=0.2) with symmetric current injection ($a_{+l}=a_{-l}=0.5$) and a nonmagnetic barrier; b) A long junction (l=2) with symmetric current injection ($a_{+l}=a_{-l}=0.5$) and a nonmagnetic barrier; c) A long (l=2) junction with symmetric current injection ($a_{+l}=a_{-l}=0.5$) and a FI barrier which is uniformly magnetized, with strength $M = 0.3M_{\text{JV}}$; d) A long (l=2) junction with asymmetric current injection ($a_{+l}=0.2,\:a_{-l}=0.8$) and a FI barrier with strength same as in (c). As can be seen, a nonreciprocal current emerges only in case (d), and this effect is present at all values of the field including zero. The red and black solid lines correspond to solutions of types $I$ and $II$, respectively, as described in the text. In Figs. (b-d), one can find multivalued solutions for $I_c$ corresponding to a given $H_e$, out of which only one is energentically favorable as discussed in Refs.~Refs.~\onlinecite{kuplevakhsky2004topological,kuplevakhsky2006static,kuplevakhsky2007exact}.}
\label{F2}
\end{figure}

It is seen that a diode effect at both zero and non-zero magnetic fields emerges only in Case 4, supporting our claim. An interesting feature is the presence of overlapping Fraunhofer lobes in a long JJ, as seen in Fig.~\ref{F2}(b,c,d). The points where $I_{c}=0$ correspond to an integral number of Josephson vortices. The central lobe corresponds to the 0-1 vortex mode, with 1 full vortex in the JJ when $I_{c}$ vanishes. The $n$ to $n+1$ vortex mode overlaps with $n+1$ to $n+2$ vortex mode since different vortex modes correspond to different self fields, resulting in different critical currents even for the same $H_{e}$ value. 

\section{Fraunhofer patterns of long and short magnetic Josephson junctions}\label{SM8}

The inset to upper panel of Fig. \ref{subfigexp4}, critical current dependence during the virgin curve of the field sequence, where the nonreciprocal effect is visible at zero field. Moreover, the nonreciprocal effect is also evident at zero field in both return branches. For comparison, we show the measured Fraunhofer pattern for the short junction in Fig.~\ref{subfigexp4}(b). No observable nonreciprocal effect is found at zero or finite fields, and the only noticeable departure from conventional Fraunhofer patterns is the appearance of hysteresis – a feature that is expected in magnetic Josephson junctions\cite{pal2014pure}.

\begin{figure}[H]
\begin{center}
\includegraphics[width=0.65\linewidth]
{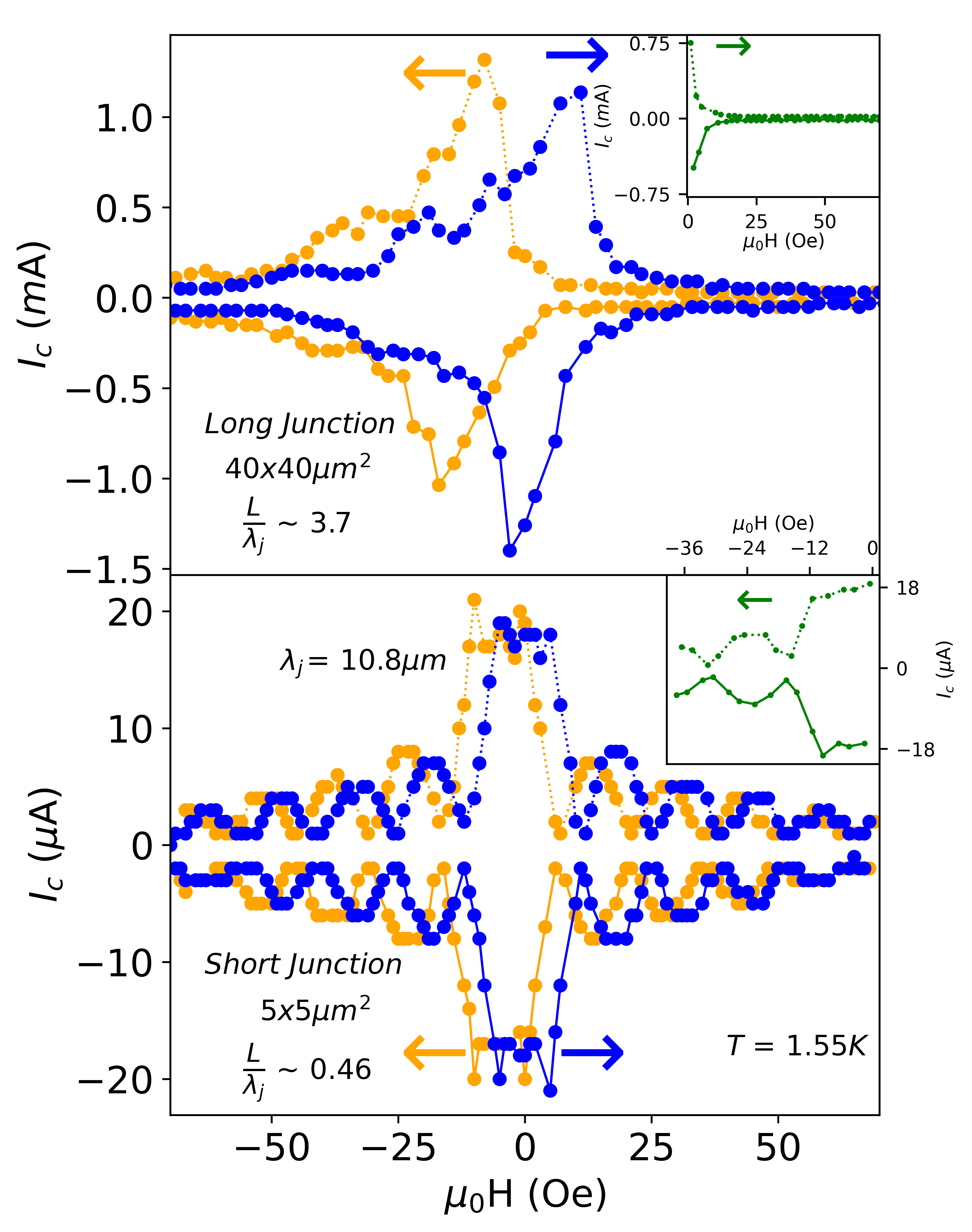}
\end{center}
\caption{Upper panel shows $I_{c}$ vs $H$ pattern for $40\:\mu \text{m} \times 40\:\mu \text{m}$ long junction. Lower panel shows $I_{c}$ vs $H$ pattern obtained from $5\:\mu \text{m} \times 5\:\mu \text{m}$ short junction. Blue (orange) markers represent the magnetic field ramp up (down) directions. Insets in both panels show $I_{c}$ vs $H$ at virgin state.} 
\label{subfigexp4}
\end{figure}

\section{Evolution of zero field efficiency with changing $L⁄\lambda_{J}$  ratio }\label{SM7}

Fig.\ref{subfigexp5} shows IV characteristic of the junctions of increasing dimension at as-cooled condition without exposing to the external field. As the ratio ($L⁄\lambda_{J}$) increases from $ $ 1 to 3, asymmetry evolves in $I_{c+}$ and $I_{c-}$, resulting in non-reciprocal Josephson effect. IV plot for ($L⁄\lambda_{J}$) $\sim 4$ is shown in the main text of the manuscript.

\begin{figure}[H]
\begin{center}
\includegraphics[width=0.7\linewidth]{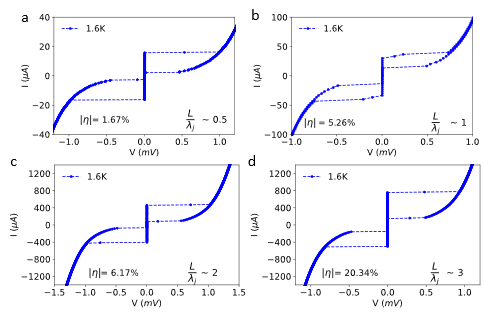}
\end{center}
\caption{IV plot for different area junctions. a) and b) IV characteristic for $5 \mu m\times 5 \mu m$ and $10 \mu m\times 10 \mu m$ respectively, showing symmetrical $I_{c+}$ and $I_{c-}$. c) and d) IV characteristic of $20 \mu m\times 20 \mu m$ and $30 \mu m\times 30 \mu m$ Junctions showing asymmetrical $I_{c+}$ and $I_{c-}$.} 
\label{subfigexp5}
\end{figure}

\section{$M-H$ loops of GdN and its usage in calculation of Josephson penetration depth} \label{SM4}
MH loops of $3nm$ thick un-patterned GdN are shown in Fig. \ref{supfigexp1} at 2K and 4.2K. The saturation moment is $\sim20 \mu emu$. Coercive field is observed at $\sim 130$Oe at 2K. As the temperature is increased coercive field is reduced to $\sim 70$Oe at 4.2K.  

\begin{figure}[H]
\begin{center}
\includegraphics[width=0.5\linewidth]{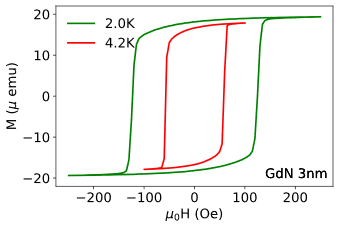}
\hspace{1mm}
\end{center}
\caption{MH plot of 3nm thick un-patterned GdN recorded at 2K and 4.2K.}
\label{supfigexp1}
\end{figure}

Magnetic permeability of GdN is $\mu_{r}=1 + \chi$, where $\chi$ is the susceptibility obtained by fitting a straight line in the M-H low field minor loops (+-10Oe). The values of $\chi$ obtained from this procedure is 0.75 at 2K. This gives us the best possible value of  $d'$ that can be used for calculation of $\lambda_{J}$. We use $\lambda_{bot}= \lambda_{top}=250nm$ for NbN. Critical current density $J_c$ of the $5\:\mu \text{m} \times 5\:\mu \text{m}$ short junction is used for the calculations. Using these values, we estimate $\lambda_J=10.8\mu \text{m}$ at 2K. Owing to temperature limitations in SQUID VSM system, we are unable to measure MH below 2K. We therefore assume that $\lambda_{J}$ values would be relatively similar at 1.55K as compared to 2K.  $\chi$ is highest in the switching regions, and from MH data shown in Figure 7, we estimate that the highest value of $\chi$ at 2K is 67. Using this high value of $\chi$, we obtain the lowest possible value of $\lambda_J=10.5\mu \text{m}$. These calculations therefore ensure that our larger junctions are very firmly in the long junction limit.

\end{appendix}
\end{widetext}

%

\end{document}